\documentclass[12pt]{article}
\usepackage{a4wide}
\usepackage{amsmath,amsthm,amsfonts}
\usepackage{graphicx}

\newtheorem{theorem}{Theorem}[section]

\newtheorem{remark}[theorem]{Remark}

\newcommand{\url}[1]{{\tt \small #1}}
\newcommand{\Ex}[2]{\mathbb{E}_{#1}\!\left[\,#2\,\right]}
\newcommand{\ExT}[3]{\mathbb{E}_{#1}^{#2}\!\left[\,#3\,\right]}
\newcommand{\Qx}[1]{\mathbb{Q}\left\{\,#1\,\right\}}

\newcommand{\ind}[1]{1_{\{#1\}}}

\title{{\large \bf Interest-Rate Modeling with Multiple Yield Curves}
\thanks
{We thank Marco Bianchetti, Andrea Germani, Fabio Mercurio, and Nicola Moreni for useful comments and hints.
}}
\author{
Andrea Pallavicini\thanks{Banca Leonardo, {\tt andrea.pallavicini@bancaleonardo.com}}
\ \ \
Marco Tarenghi\thanks{Mediobanca, {\tt marco.tarenghi@mediobanca.it}}
}
\date{\small First Version: October 13, 2009.  This version: \today}

\begin{document}

\maketitle

\begin{abstract}
The crisis that affected financial markets in the last years leaded market practitioners to revise well known basic concepts like the ones of discount factors and forward rates. A single yield curve is not sufficient any longer to describe the market of interest rate products. On the other hand, using different yield curves at the same time requires a reformulation of most of the basic assumptions made in interest rate models. In this paper we discuss market evidences that led to the introduction of a series of different yield curves. We then define a HJM framework based on a multi-curve approach, presenting also a bootstrapping algorithm used to fit these different yield curves to market prices of plain-vanilla contracts such as basic Interest Rate Swaps (IRS) and Forward Rate Agreements (FRA). We then show how our approach can be used in practice when pricing other interest rate products, such as forward starting IRS, plain-vanilla European Swaptions, Constant Maturity Swaps (CMS) and CMS spread options, with the final goal to investigate whether the market is actually using a multi-curve approach or not. We finally present some numerical examples for a simple formulation of the framework which embeds by construction the multi-curve structure; once the model is calibrated to market prices of plain-vanilla options, it can be used via a Monte Carlo simulation to price more complicated exotic options.

\end{abstract}

{\bf JEL classification code: G13. \\ \indent AMS classification codes: 60J75, 91B70}

\medskip


\newpage
{\small \tableofcontents}
\vfill
{\footnotesize \noindent The opinions expressed here are solely those of the authors and do not represent in any way those of their employers.}
\newpage

\section{Introduction}

Classical interest-rate models are formulated to embed by construction non-arbitrage relationships, which allow to hedge forward-rate agreements in terms of zero-coupon bonds. As a direct consequence, models predict that forward rates of different tenors are related to each other by sharp constraints; however these non-arbitrage relationships might not hold in practice (think for example of basis-swap spreads which, from a theoretical point of view should be equal to zero, but they are actually traded in the market at quotes larger than zero). 

Then, the market virtually presents situations of possible arbitrage violations. This is what happened starting from summer 2007, with the raising of the credit crunch, where market quotes of forward rates and zero-coupon bonds began to violate the usual non-arbitrage relationships in a macroscopic way, both under the pressure of a liquidity crisis, which reduced the credit lines needed to hedge unfunded products using zero-coupon bonds, and the possibility of a systemic break-down suggesting that counterparty risk cannot be considered negligible any more. The resulting picture, as suggested by Henrard (2007), describes a money market where each forward rate seems to act as a different underlying asset.

\begin{figure}[htp]
\begin{center}
\includegraphics[scale=0.5]{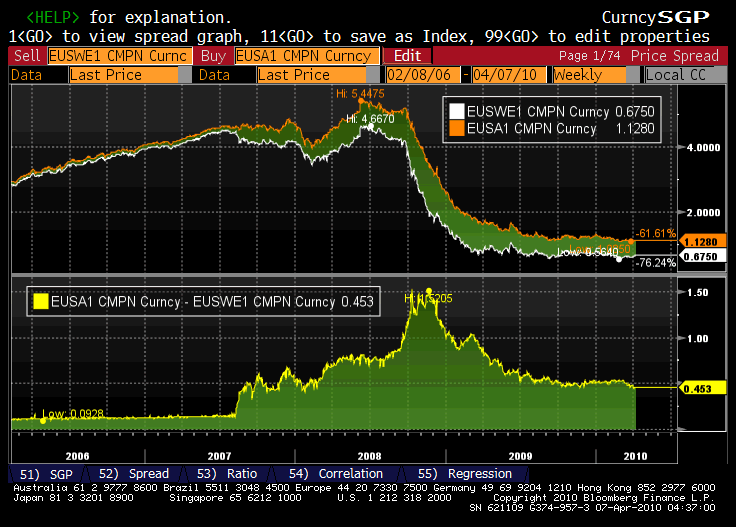}

\vspace*{1cm}
\includegraphics[scale=0.5]{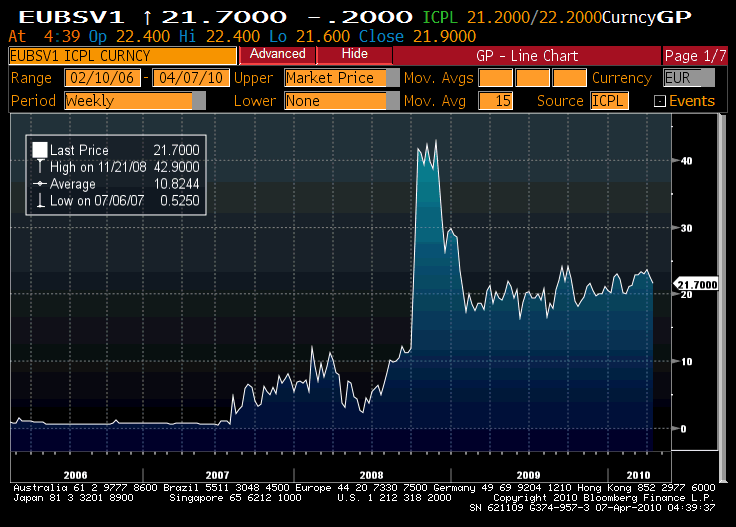}
\end{center}
\caption{
\label{fig:bloom}
Upper panel: EONIA swap rate with maturity one year (white line) and EURIBOR swap-rate (orange line) on the same maturity. Lower panel: basis-swap spread for six-months vs. three-months EURIBOR rates on a swap with maturity one year. The two figures are obtained from Bloomberg\textsuperscript{\textregistered} platform.
}
\end{figure}

In the upper panel of Figure \ref{fig:bloom} we show, starting from 2006, the weekly history of the EONIA swap rates with maturity one year and EURIBOR swap rates on the same maturity. The spread between the two rates may be considered as an indicator of the presence of a growing systemic risk in the Euro area as soon as the crisis breaks out. In the lower panel we show the basis-swap spread for six-months vs. three-months EURIBOR rates on a swap with one year maturity over the same temporal scale. In both cases we can observe a similar behaviour: before summer 2007 the spread between Overnight Indexed Swaps (OIS) and standard Interest Rate Swaps (IRS) was quite low and constant, indicating low liquidity and counterparty risk, and basis-swap spreads were nearly zero, consistently with the usual interest-rate models predictions. With the beginning of the crisis, this situation changed abruptly: spreads between OIS and IRS widened, and traded basis-swap spreads are now significantly different from zero. In particular the lower panel of Figure \ref{fig:bloom} shows that during the autumn 2008 the market value of basis-swap spread was about $40$ basis points, which is fully inconsistent with traditional interest-rate models. In such situations a full-featured model comprehensive of both liquidity and credit risks should be used, but it is far from being forged and ready to be used by practitioners, so that, in order to overcome the problems posed by using a unique yield curve for all forward rates, many simplified approaches were developed by single banks or institutions.

The simplest receipt consists in using a different yield curve for each product, namely if the instrument's underlying is a particular rate with a given tenor, then a yield curve, bootstrapped only from quoted instruments based on such underlying, is employed both to discount cash flows and to calculate index fixings. This approach works for simple products, but it fails, for instance, as soon as we consider products depending on forward rates of different tenors.

Recently in the literature some authors started to deal with these issues, mainly concerning the valuation of cross currency swaps as in Fruchard {\it et al.}~(1995), Boenkost and Schmidt (2005), Kijima {\it et al.}~(2009), and Henrard~(2007, 2009). They face the problem in a pragmatic way by considering each forward rate as a single asset without investigating the complex dynamics involved by liquidity and credit risks, although a temptatives are made in Morini (2009) and Morini and Prampolini (2010). In particular they propose methods to extend yield curve bootstrapping to a multi-curve setting, as in Chibane and Sheldon~(2009), in Bianchetti~(2009) or in Kenyon~(2010), or new pricing models as in Kijima {\it et al.}~(2009) or in Mercurio~(2009). We cite also a slightly different approach by Fujii {\it et al.}~(2010) and Mercurio~(2010), where they explicitly model each basis spread.

In this paper we start in section \ref{sec:hjm} by introducing a HJM framework which is able to incorporate all the initial yield curves, once a discounting curve is given, without assuming the usual non-arbitrage conditions on interest-rates. In section \ref{sec:bootstrap} we present an efficient method for bootstrapping the initial yield curves starting from market quotes of basic linear interest rate instruments, such as Interest Rate Swaps (IRS) and Forward Rate Agreements (FRA). In section \ref{sec:evidence} we analyze the pricing of some more sophisticated instruments, like forward starting IRS, plain-vanilla European swaptions, Constant Maturity Swaps (CMS) and CMS spread options, trying to understand whether the market is actually using the multi-curve approach or not. In particular we will provide evidences that there is a market segmentation between basic money market instruments (spot and forward IRS, FRA, etc.), that seem to be priced using the multi-curve setting, and interest rate options, that are still priced using a single-curve setting. Then, in section \ref{sec:gaussian} we define a restricted version of the model and we calibrate it to plain-vanilla swaptions. Finally, a section of conclusions closes the paper.%Exotic pricing in section \ref{sec:exotics} and conclusions end the paper.

%Possible discounting curves may be either the Overnight Indexed Swap (OIS) curve (for collateralized derivatives), the LIBOR deposit curve (to reflect bank systemic risk) or, for specific counterparties, it can be constructed by imposing a credit spread over the risk-free curve, whose size is calculated according to the bilateral credit valuation adjustment. The paper continues with numerical examples of the techinques used to bootstrap the yield curves, based on modelling the spread between the curves, and with a joint calibration of the HJM model to cap and swaption volatilities. It concludes with some examples of exotic option pricing.

\section{HJM models with independent forward rates}
\label{sec:hjm}

Interest-rate derivatives are actively traded in the market. Many plain-vanilla contracts are present in the market and their prices are quite liquid. It is a common practice to start by bootstrapping an initial yield curve from market quotes of deposits, Forward Rate Agreements (FRA), short futures and standard Interest Rate Swaps (IRS), and, as a second step, to introduce an interest-rate model able to incorporate such initial term-structure with enough free parameters to calibrate interest rate derivatives such as caps and swaptions.

However, in the last three years, as the credit crisis grew, many problems arose in interpreting all these quotes by means of a unique term-structure. The well known non-arbitrage relationship between forward rates ($F$) ans zero-coupon bonds ($P$) cannot be used safely any longer, namely the equality
\[
F_t(T_0,T_1) = \frac{1}{T_1-T_0}\left(\frac{P_t(T_0)}{P_t(T_1)}-1\right)
\]
may be violated due to the presence of liquidity and counterparty risk, so that we are tempted to introduce different term-structures: one for the discounting curve, and many for the forwarding curves, one for each interest-rate tenor, to recover a realistic description of market quotes.

Indeed, if we try to replicate a forward rate by means of zero-coupon bonds, we are forced both to buy and sell a bond. During the crisis many banks closed their credit lines with counterparties resulting in the impossibility for trading desks to effectively build such replication strategy, so that prices of zero-coupon bonds and forward contracts, not being any longer constrained, may depart from the values required by non-arbitrage. In a more general context, a strong distinction between \textit{funded} and \textit{unfunded} products is needed: because of the presence of liquidity and credit issues, basic funded products like simple deposits cannot be used to infer a risk-free discounting curve. On the other hand, unfunded and collateralized products like IRS have cash-flows depending on rates like EURIBOR which embed by their definition a credit risk.

We can add counterparty risk to the picture, by considering the possibility of a systemic default. If we consider that forward rates are defined as the fair values of FRA contracts, we can write following Mercurio (2009):
\[
\Pi_{\rm FRA}(t,T_0,T_1;K) := \Ex{t}{\frac{(T_1-T_0)(K-L_{T_0}(T_1))}{1+(T_1-T_0) L_{T_0}(T_1)}\frac{B_t}{B_{T_0}}}
\]
where $L_{T_0}(T_1)$ is the Libor rate fixing at time $T_0$ and paying at time $T_1$, and $B_t$ is the bank account process. If the contract is sold at par we can solve for $K$, and we get
\begin{eqnarray*}
K = \frac{1}{T_1-T_0} \left( \frac{P_t(T_0)}{P_t(T_1)Q_t(T_0,T_1)} - 1 \right)
\end{eqnarray*}
where $Q_t(T_0,T_1)$ is defined as
\[
Q_t(T_0,T_1) := \frac{1}{P_t(T_1)} \,\Ex{t}{\frac{1}{1+(T_1-T_0) L_{T_0}(T_1)}\frac{B_t}{B_{T_0}}}
\]

Then, since the Libor rate $L$ is the reference lending rate, we can define it in term of the risky bond ${\bar P}_t(T)$ as
\[
L_{T_0}(T_1) := \frac{1}{T_1-T_0}\left(\frac{1}{{\bar P}_{T_0}(T_1)}-1\right)
\;,\quad
{\bar P}_t(T) := \Ex{t}{\frac{B_t}{B_T}\ind{\tau>T}\Big|\tau>t}
\]
where $\tau$ is the systemic default event. We can plug the formula into the definition of $Q$:
\[
Q_t(T_0,T_1) = \frac{1}{P_t(T_1)} \,\Ex{t}{{\bar P}_{T_0}(T_1)\frac{B_t}{B_{T_0}}}
\]

For sake of sketching the effect of counterparty risk on forward rates we consider that the default probability is independent from the interest rates, leading to
\[
{\bar P}_t(T) = P_t(T) \Qx{\tau>T|\tau>t}
\;,\quad
Q_t(T_0,T_1) = \Qx{\tau>T_1|\tau>T_0}
\]
Notice that according to the above formula we get that forward rates with different tenors belong to different term-structures, and, in particular, since the higher is the tenor the lesser is the conditional survival probability $Q$, we get  positive basis relating lower tenors to higher tenors, as quoted by the market. See also Morini (2009).

However, as done by many authors in the literature, starting from Henrard (2007), we avoid to directly model liquidity or counterparty risk but we follow a more pragmatic line. We do not try to model why the yield curves differ, but we describe how to consider different dynamics for forward rates of different tenors.

\subsection{The multi-curve framework}

We start by introducing a different HJM model for each different Libor rate tenor. For a standard introduction to the HJM framework see Brigo and Mercurio (2006), but see also Liu and Wu (2008) where inflation dynamics is modeled within a two-curve HJM framework.

Let us define a different family of  instantaneous forward rates $f^\Delta_t$ for each Libor rate with tenor $\Delta$, whose dynamics under risk-neutral measure is given by
\[
df^\Delta_t(T) = \sigma^\Delta_t(T) \left( \int_t^T du \, \sigma^\Delta_t(u) \right) \,dt + \sigma^\Delta_t(T) \,dW^\Delta_t
\]
where $\sigma^\Delta_t(T)$ is a volatility process. Notice that the above expression can be readily generalized to consider a vector of Brownian motions.

Zero-coupon bonds $P_t^\Delta(T)$ can be expressed in term of instantaneous forward rates by means of
\[
P_t^\Delta(T) := \exp\left\{-\int_t^T du \, f^\Delta_t(u)\right\}
\]
from which we can derive the zero-coupon bond dynamics under the risk-neutral measure
\[
\frac{dP_t^\Delta(T)}{P_t^\Delta(T)} = r^\Delta_t \,dt - \left( \int_t^T du \, \sigma^\Delta_t(u) \right) \, dW^\Delta_t 
\]
where the drift terms $r_t^\Delta$ are generally different from the short rate $r_t$ used for discounting, to reflect the fact that Libor rates are not any longer linked to discounting zero-coupon bonds.

Indeed, since the HJM model requires the non-arbitrage relationship linking Libor rates and zero-coupon bonds to work, we simply state that the zero-coupon bonds appearing in such HJM models are different from the ones we use for discounting.
\[
L_{T-\Delta}(T) = \frac{1}{\Delta} \left( \frac{1}{P^\Delta_{T-\Delta}(T)} - 1 \right) \neq \frac{1}{\Delta} \left( \frac{1}{P_{T-\Delta}(T)} - 1 \right)
\]
where $\Delta$ is the Libor rate tenor.

In order to link Libor rates to market quotes, we have to price contracts paying the Libor rate such as
\[
\Pi(t,T-\Delta,T) := \Ex{t}{ \Delta L_{T-\Delta}(T) \frac{B_t}{B_{T}} } 
\]
Thus, according to the usual HJM framework we should introduce a family of forward rates, starting from zero-coupon bond definition%, and we should relate them to the expectations of the Libor rates under forward measure, namely
\[
F_t(T-\Delta,T) := \frac{1}{\Delta} \left( \frac{P^\Delta_t(T-\Delta)}{P^\Delta_t(T)} - 1 \right)
%\;,\quad
%F_t(T-\Delta,T) = \ExT{t}{P^\Delta(T)}{L_{T-\Delta}(T)}
\]
%Notice that the expectation has to be taken under the $P^\Delta(T)$-forward measure and not under the $P(T)$-forward measure (or discounting forward measure). 
On the other hand, in the multi-curve HJM framework, if we consider a contract paying the Libor rate we get% a very different pricing formula
\[
\Pi(t,T-\Delta,T) = \Ex{t}{ \Delta L_{T-\Delta}(T) \frac{B_t}{B_{T}} } \neq \Delta F_t(T-\Delta,T) P_t(T)
\]

Hence, in the multi-curve HJM framework it is useful to define the following modified forward rate (see also Mercurio (2009) for an analogous definition) under the discounting forward measure, which enters most of the pricing formulas of contracts quoted by the market.
\begin{equation}
\label{modfwd}
\tilde{F}_t(T-\Delta,T) := \ExT{t}{P(T)}{L_{T-\Delta}(T)}
\end{equation}
Notice that the modified forward rate $\tilde{F}$ is a martingale by construction under the discounting forward measure. Further, we get
\[
\Pi(t,T-\Delta,T) = \Delta \tilde{F}_t(T-\Delta,T) P_t(T)
\]

Forward-rate dynamics can be derived from zero-coupon bond dynamics by means of It\^o formula keeping into account that forward rates are not martingales under the discounting forward measure. We get, under risk-neutral measure
\begin{equation}
\frac{dP_t(T)}{P_t(T)} = r_t \,dt - \left( \int_t^T du \, \sigma_t(u) \right) \, dW_t
\end{equation}
and, under discounting forward measure
\begin{equation}
\frac{dF_t(T-\Delta,T)}{F_t(T-\Delta,T) + \Delta^{-1}} = \left(\int_{T-\Delta}^{T} du \,\sigma^\Delta_t(u) \right) \left( \vartheta^\Delta_t(T) \,dt + \,dW^\Delta_t \right)
\end{equation}
with
\[
\vartheta^\Delta_t(T) := \int_t^T du \left( \sigma^\Delta_t(u)-\rho^\Delta\sigma_t(u) \right)
\]
and
\[
\rho^\Delta := \frac{d}{dt} \langle W,W^\Delta \rangle_t
\]

Then, by direct integration it is possible to calculate the ${\tilde F}$s as given by
\begin{equation}
{\tilde F}_t(T-\Delta,T) = F_t(T-\Delta,T) \left( 1 + \frac{1+\Delta F_t(T-\Delta,T)}{\Delta F_t(T-\Delta,T)}(\Theta_t^\Delta(T)-1) \right)
\end{equation}
where
\[
\Theta_t^\Delta(T) := \exp\left\{ \int_t^{T-\Delta} du \int_{T-\Delta}^{T} dv \,\sigma^\Delta_u(v) \,\vartheta^\Delta_u(T) \right\}
\]

\subsection{Discounting and funding curves}

In the multi-curve HJM framework, risk-neutral measure will be defined as usual with respect to the bank account used for risk-free funding. Since risk-free funding can be done in term of overnight rates\footnote{We assume here that the credit risk embedded in an overnight loan is negligible.}, we get
\[
B_t := \prod_{i=1}^{n(t)} \left( 1+(t_i-t_{i-1}) c_{t_i} \right)
\]
where $c_t$ is the overnight rate at time $t$, and $n(t)$ is the number of business days between time $0$ and time $t$. In the Euro area the discounting curve can be obtained starting from the EONIA-based Overnight Indexed Swaps, that typically cover a time horizon up to thirty years. The use of a discounting curve obtained from overnight rates is a typical choice in the multi-curve setting, and it is made, among the others, by Fujii {\it et al.}~(2010) and Mercurio~(2010). This choice can be justified by the the fact that, usually, interbank operations are collateralized; if we assume that the collateral is revalued daily, then it is straightforward to use an overnight rate (such as the EONIA rate) for discounting. As noticed by Henrard~(2009), different choices for the discounting curve can lead to different prices of interest rate derivatives, and even if the difference is not relevant, neverthless it is not negligible. In section~\ref{sec:evidence} we present some evidences of market quotes of Interest Rate Swaps that are coherent with the choice of a discounting curve obtained from overnight rates.

From the bank account we can then define a discounting short rate $r_t$ and zero-coupon bonds $P_t(T)$ by the usual formula and model them via a HJM model as given in the previous section.

\begin{remark} {\bf Counterparty and liquidity risk:}
If funding cannot be done in a risk-free way, we should keep into account the possibility of default of the counterparties entering the deal, and as a consequence we cannot define a unique funding curve, but it does depend on the credit worthiness of the counterparties.

A possible approach is keeping a risk-free curve for discounting, but adjusting the pricing formula to include bilateral counterparty risk, whose definition can be found in Brigo and Capponi (2008), and also liquidity risk as done in Pallavicini (2010) although in a rough way.

Notice that also collateralization impacts the definition of the funding curve, see for instance Piterbarg (2010).
\end{remark}

\section{Bootstrapping the initial yield curves}
\label{sec:bootstrap}

In our HJM multi-curve framework we have one risk-free discounting curve and many forwarding curves, one for each quoted Libor rate tenor:
\[
T \mapsto P_t(T) \;,\quad T \mapsto {\tilde F}_t(T-\Delta,T) \,,\; \Delta\in{{\rm 1m},{\rm 3m},{\rm 6m},{\rm 12m}}
\]
As a practical example we consider the Euro area. 

The problem of bootstrapping different curves corresponding to rates of different tenors has been addressed for example in Ametrano and Bianchetti~(2009) where they solve the problem in terms of different yield-curves coherent with market quotes of basic derivatives, or in Bianchetti~(2009) where they follow an approach based on the FX equivalence to discriminate between different curves as if they were curves corresponding to different currencies.

Here, we follow an independent approach to bootstrap the term structures. We start by considering the spread between modified forward rates of different tenor, see equation \eqref{modfwd}, along with their spread with respect to forward rates calculated from the discounting curve. We bootstrap the yield curves by interpolating on such spreads. In particular our aim is to produce smooth curves of forward rates and basis spreads

% The zero-coupon bond curves $P^{\Delta}_t(T)$ are not direct market observables, but on the contrary we can observe the modified forward rates $\tilde{F}_t(T-\Delta,T)$ defined in (\ref{modfwd}). This is the consideration that leads to the direct modeling of such rates in Mercurio~(2009), or of the spread between them and the forward rate obtained from the discounting curve as in Fujii {\it et al.}~(2010) and Mercurio~(2010). We therefore concentrate directly on bootstrapping term structures of these modified forward rates which are coherent with market quotes of interest rates basic derivatives, and that could be directly interpolated at each date in the future. In particular, this last feature overcomes the interpolation problems found by Ametrano and Bianchetti~(2009), where apparently smooth interpolation choices for the discount curve lead to unstable forward rates: here, in fact, we directly assume an interpolation structure on the forward rates.

Our bootstrap procedure is based on the assumption of a given discounting curve, which in the following we refer to also as the 1d-tenor curve. As hinted at in the previous section, we calibrate it (using standard techniques) against the following instruments:
\begin{itemize}
\item {\bf Discounting:} EONIA fixing, OIS from one to thirty years,
\end{itemize}

Once the discounting curve is known, the forwarding curves are calibrated to the following market quotes, listed below according to underlying tenors:
\begin{itemize}
\item {\bf Indexation over 1m:} EURIBOR one-month fixing, swaps from one to thirty years paying an annual fix rate in exchange for the EURIBOR 1m rate (some of these swaps may be substituted with one-vs-three-months basis-swaps);
\item {\bf Indexation over 3m:} EURIBOR three-months fixing, Short Futures, FRA rates up to one year, swaps from one to thirty years paying an annual fix rate in exchange for the EURIBOR 3m rate (some of these swaps may be substituted with at-the-money cap strikes or with three-vs-six-months basis-swaps);
\item {\bf Indexation over 6m:} EURIBOR six-months fixing, FRA rates up to one year and a half, swaps from one to thirty years paying an annual fix rate in exchange for the EURIBOR 6m rate (some of these swaps may be substituted with at-the-money cap strikes);
\item {\bf Indexation over 12m:} EURIBOR twelve-months fixing, FRA rates up to two years, six-vs-twelve-months basis-swaps from two to thirty years.
\end{itemize}

Notice that the payoffs of these instruments must be calculated without resorting to the usual non-arbitrage relationships. In particular for FRA, IRS and Basis Swaps we get relevant modifications, see also Mercurio (2009).

Before entering the details of the bootstrap procedure, let us give a brief look to the main interest rate instruments considered.

\subsection{Libor fixings}
According to formula (\ref{modfwd}) we can set the value of the modified forward at spot date\footnote{For Euro area spot date is two business days after trade date, but for sake of simplicity we set $t=0$ in out formula, although numerical calculations are always performed taking into account all market conventions.}, namely $\tilde{F}_0(0,\Delta)$, equal to the fixing of the Euribor rate corresponding to the given tenor~$\Delta$.

\subsection{FRA par-rates}

The par rate $K$ of FRA contracts is modified by a convexity adjustment. Indeed, we get
\begin{eqnarray}
K &=& \frac{\Ex{0}{\frac{F_t(t,T)}{1+(T-t) F_t(t,T)}D(0,t)}}{\Ex{0}{\frac{1}{1+(T-t) F_t(t,T)}D(0,t)}}
   =  \frac{\Ex{0}{(1-P^{T-t}_t(T))D(0,t)}}{(T-t)\Ex{0}{P^{T-t}_t(T)D(0,t)}} \nonumber \\
  &=& \frac{1}{T-t} \left( \frac{1}{\ExT{0}{t}{P^{T-t}_t(T)}} - 1 \right)
\end{eqnarray}
which cannot be defined in term of the $\tilde F$s. Thus, FRAs in a multi-curve framework need a correction like futures.

\subsection{Swap par-rates and basis-swap spreads}

We can recalculate market quotes for any swap in term of forward rates ${\tilde F}_t(T-\Delta,T)$ for different tenors $\Delta$. Indeed, for a standard IRS we get that the fair rate $S$ is given by:
\begin{equation}
S_0(0,T_n) := \frac{\Delta \sum_{i=1}^n {\tilde F}_0(T_i-\Delta,T_i) P_0(T_i)}{\Delta' \sum_{j=1}^{n'} P_0(T'_j)}
\end{equation}\label{fairirs}
where $T_1,\ldots,T_n$ are the payment dates of the floating leg, $T'_1,\ldots,T'_{n'}$ are the payment dates of the fixing leg. The definition of the forward-starting swap rate $S_0(T_a,T_b)$ follows accordingly.

Analogously, for a basis-swap, the fair spread $X$ is given by:
\[
X := \frac{\Delta \sum_{i=1}^n {\tilde F}_0(T_i-\Delta,T_i) P_0(T_i) - \Delta' \sum_{j=1}^{n'} {\tilde F}_0(T'_j-\Delta,T'_j) P_0(T'_j)}{\Delta' \sum_{j=1}^{n'} P_0(T'_j)}
\]
which is generally different from zero.

\subsection{The bootstrapping procedure}

We start by considering the risk-free discounting curve. Since the quotes for EONIA and OIS depend only on this curve, we can bootstrap it by means of the usual techniques, for instance we choose to employ the monotone cubic interpolation based on Hermite polynomials (see Hagan and West (2006) and Ametrano and Bianchetti (2009) for a review of bootstrapping techniques).

Once the discounting curve is known, we derive from it a curve of 1d forward rates obtained by using the standard no arbitrage arguments: 
\[
\tilde{F}_0(t-\Delta_{1d},t) \equiv F_0(t-\Delta_{1d},t) = \frac{1}{\Delta_{1d}}\left(\frac{P_0(t-\Delta_{1d})}{P_0(t)}-1\right)
\]
then, starting from this curve, we obtain the forwarding curves corresponding to the different rate tenors. 

We start from the six-months tenor, which corresponds in the Euro area to the family of most liquid instruments. We take the following steps:
\begin{enumerate}
\item we define the rate difference $y_{\rm 6m/1d}(t) := {\tilde F}_0(t-\Delta_{\rm 6m},t) - {\tilde F}_0(t-\Delta_{\rm 1d},t)$ (remember that we use the label 1d to refer to the discounting curve);
\item we bootstrap the curve of the $y$s to match the six-months-tenor market quotes by using as interpolation scheme the monotone cubic interpolation based on Hermite polynomials;
\item we get the curve of the six-months $\tilde F$s by inverting the definition of the $y$s.
\end{enumerate}

Notice that we choose to bootstrap the rate differences $y$, instead of directly acting on the rates $\tilde F$, so that the interpolation scheme can produce a smoother basis between the six-months and the 1-day forward rates.

Once we know the six-months curve, we can proceed in a similar way (interpolation on rate differences) to obtain the curves corresponding to the other tenors. We consider the liquidity of the underlying instruments to select which rate difference we want to bootstrap, and
\begin{enumerate}
\setcounter{enumi}{3}
\item we obtain the three-months curve using as starting point the six-months curve, since the market quotes the three-vs-six-months basis-swaps;
\item we obtain the one-month curve using as starting point the three-months curve, since the market quotes the one-vs-three-months basis-swaps;
\item we obtain the twelve-months curve using as reference the six-months curve, since the market quotes the six-vs-twelve-months basis-swaps.
\end{enumerate}

\begin{figure}
\begin{center}
\includegraphics[scale=0.5]{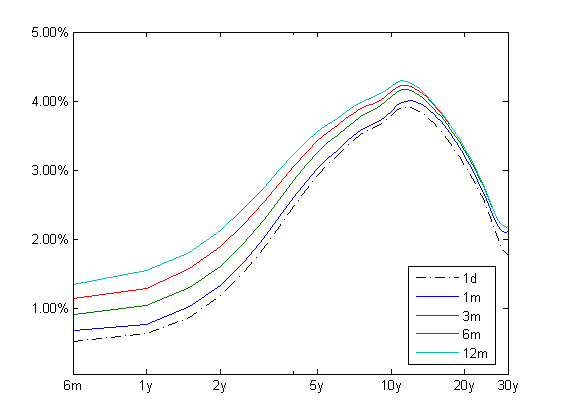}
\hfill
\includegraphics[scale=0.5]{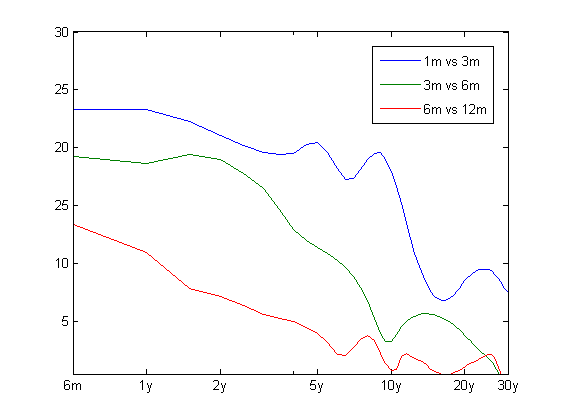}
\end{center}
\caption{
\label{fig:hjm}
Curve bootstrap in a multi-curve framework with monotone cubic interpolation based on Hermite polynomials. Left panel shows forward rates of different tenors, where 1d means EONIA (on the $x$-axis we have the rate forward start date expressed in years from the evaluation date while on the $y$-axis we have the value of the rate). Right panel forward basis-swap spreads for a 1y-length swap (on the $x$-axis we have the swap forward start date while on the $y$-axis we have the value of the spread in basis points). Market data observed on 14 June 2010.
}
\end{figure}

The results of this procedure are shown in Figure \ref{fig:hjm}. Notice that all the different curves present a smooth behaviour over time; moreover, they respect the expected fact that larger tenors should correspond to larger rates, indicating a larger impact of credit/liquidity issues.

%\begin{remark} {\bf Exact calibration vs. bestfit:}
%Classical discount curves (in the single curve framework) are generally obtained from an exact calibration to the market quotes chosen for the bootstrap: this is possible since one pillar value for the discounting curve is chosen for each pillar instrument considered. If we try and apply a similar procedure to the calibration of the forward rates, we find very unstable results; so we decide to consider a number of free parameters smaller than the number of the calibrated instruments (tipically associated to the most liquid maturities), and we run a minimization procedure aiming to obtain a bestfit. In this way we do not find an exact calibration to all the instruments, but the calibration errors are very small and the resulting curve is much smoother and regular.
%\end{remark}

\section{Market evidence of multi-curve pricing}
\label{sec:evidence}

Independence of forward rates is assumed to deal with basis-swap spreads consistently different from zero. However, such evidence is not unique. In this section we are going to present a survey of typical interest rate instruments available in the market, and we analyze if their quoted prices/rates are compatible with the multi-curve framework illustrated above.

For our test we get market quotes from ICAP\textsuperscript{\textregistered} pages listed on Bloomberg\textsuperscript{\textregistered} platform on June 14th, 2010.

%In particular, market quotes of forward swap rates are sensitive to possible discrepancies between the discounting and the forwarding term structures. Further, option pricing is affected by the independence of forward rates too, as we can deduce from market quotes of CMS related products, such as CMS swaps and spread options. We check in the following sections if there is on the market an evidence of derivative pricing with the multi-curve framework. We will find that the swap market seems to employ such framework to quote par rates, while a direct evidence is lacking for the CMS option market. Such results are in agreement with the anecdotical arguments found in Bianchetti (2010).

\subsection{Forward swap rates}

The money market quotes both spot and forward starting swap rates. These data are useful to check whether market practice is coherent with a multi-curve framework when pricing simple derivative contracts.

If we look at formula (\ref{fairirs}), which can be easily extended to forward starting swaps, we notice that the fair rate of an IRS depends both on the discounting curve and on the indexation curve corresponding to the floating rate tenor. Hence, the choice of these two curves becomes crucial to obtain swap rates coherent with market quotes.

\begin{figure}
\begin{center}
\includegraphics[scale=0.5]{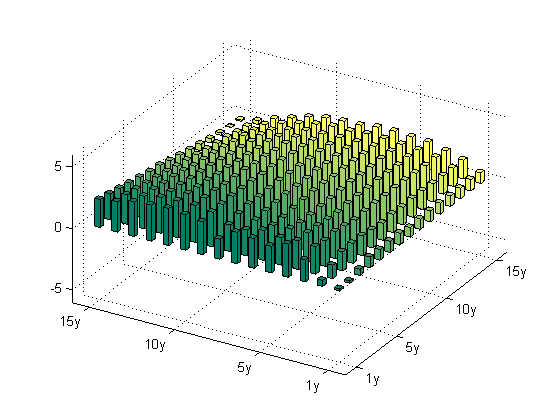}
\hfill
\includegraphics[scale=0.5]{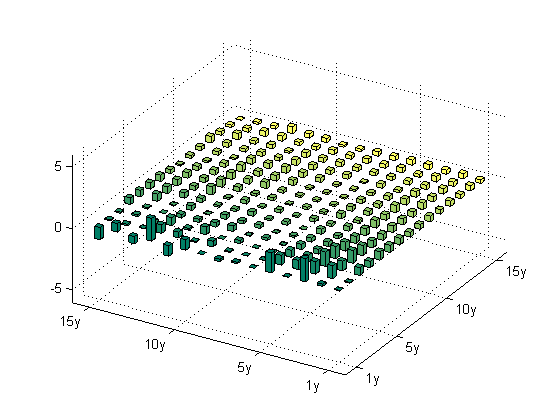}
\end{center}
\caption{
\label{fig:irsfwd}
Forward swap rate calculated in a single-curve framework (left panel) and in a multi-curve framework (right panel). Each bar represents the difference in basis points with respect to market quotes. Left axis is swap forward start date, right axis is swap tenor. Market data observed on 14 June 2010.
}
\end{figure}

We compare in Figure \ref{fig:irsfwd} a set of forward swap rates quoted by the market with the same rates calculated both in a single-curve framework\footnote{Notice that since the underlying rate tenor for IRS in the Euro market is equal to six-months, the unique curve used both for discounting and forwarding is obtained from classical bootstrap of instruments based on six-months tenor rates. This is the curve which we will refer to also in the following when we talk of single-curve pricing.} (left panel) and in a multi-curve framework (right panel). We notice that in the single-curve framework the calculated swap rates are badly fitted to market quotes while, on the other side, the swap rates calculated in the multi-curve framework are in a very good accordance with all the quoted values considered: this is a strong evidence that the market has abandoned the single-curve approach for the multi-curve approach (at least for swap rates). Also, it is a confirmation that the market is using a discounting curve obtained from instruments based on the overnight rate: this is coherent with the fact that standard IRS are collateralized instruments.

\subsection{Plain Vanilla Swaptions}

Given the accordance between the forward swap rates quoted in the market and those computed via multi-curve pricing, it is natural to check what happens for options having has underlying the forward swap rate. We can write the (payer) swaption price as:

\begin{figure}[t]
\begin{center}
\begin{tabular}{cc}
\includegraphics[scale=0.5]{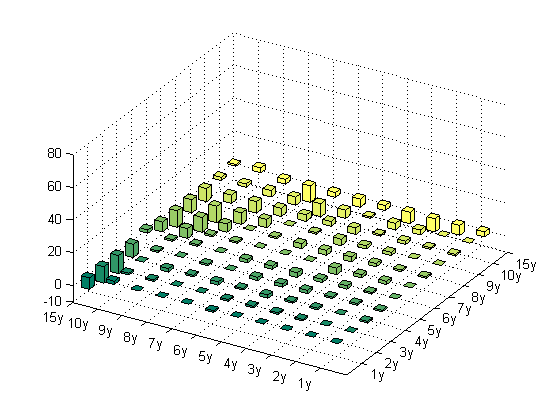} &
\includegraphics[scale=0.5]{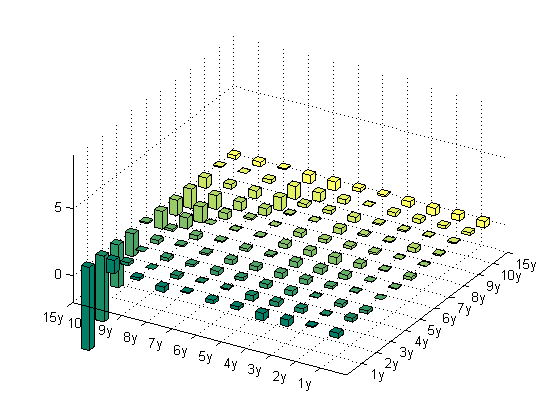} \\
\includegraphics[scale=0.5]{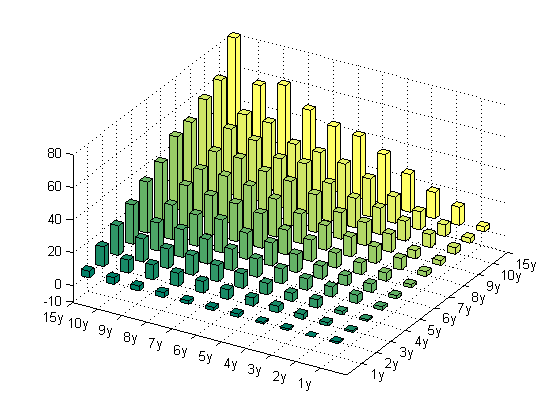} &
\includegraphics[scale=0.5]{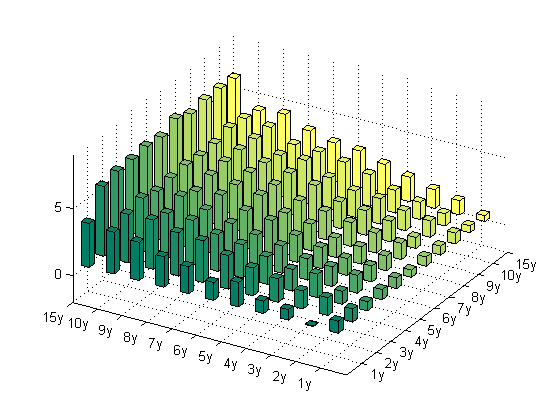}
\end{tabular}
\end{center}
\caption{
\label{fig:swaption}
Differences between market prices and prices computed starting from quoted volatility of ATM swaption straddles. In the upper panels we plot the differences (absolute errors in basis points on the left and relative errors in percentage on the right) when prices are computed in a single-curve approach while in the lower panels we plot the differences when prices are computed using a multi-curve approach. Left axis is option expiry, right axis is underlying swap tenor. Market data observed on 14 June 2010.
}
\end{figure}

\begin{equation}
\label{swaption}
\pi_0(T_a,T_b;K) = P_0(T_a)\,C_0(T_a,T_b)\,{\rm Bl}(S_0(T_a,T_b),K,v(K,T_a))
\end{equation}
where ${\rm Bl}(S,K,v)$ is the core of the Black formula for swaptions, with $S$ the forward swap rate, $K$ the strike rate, and $v$ the variance of the swap rate, and it is given by
\[
{\rm Bl}(S,K,v) := S \Phi(d_1) - K \Phi(d_2)
\;,\quad
d_1 := \frac{\ln(S/K)}{\sqrt{v}} + \frac{1}{2}\sqrt{v}
\;,\quad
d_2 := d_1 - \sqrt{v}
\]
where the variance $v=v(K,T)=\sigma^2(K)(T-T_0)$ can be obtained from quoted implied volatilities. $C_0(T_a,T_b)$ is the swaption pseudo-numeraire that can be expressed as
\[
C_0(T_a,T_b) := \sum_{i=a+1}^b \left(\frac{1}{1+(T_i-T_{i-1})S_0(T_a,T_b)}\right)^{(T_i-T_a)}
\]
for cash settled options, while for physically settled options\footnote{We recall that if the option at expiry is in-the-money, then a physically settled swaption implies that the two counterparties enter a swap contract, while a cash settled swaptions implies the payment of the discounted cash flows of the theoretical underlying swap, where (by contract) the flows are discounted using the swap rate fixing at option expiry itself. Usually, in the Euro market swaptions are traded cash.} it is replaced by
\[
\frac{1}{P_0(T_a)}\sum_{i=a+1}^b (T_i-T_{i-1}) P_0(T_i)
\]

A typical market practice is to quote plain-vanilla European swaptions according to their implied volatility coherent with the price at which they are actually traded. We computed swaption prices starting from the quoted implied volatility, both in the single and in the multi-curve framework, and we compared it with the quoted price. We report the results in Figure \ref{fig:swaption}, where we can see that market prices are better fitted when the swaptions are priced using a single-curve approach. So, even if market forward swap rates are priced in a multi-curve setting, this does not hold for European swaptions, that are still priced using a single-curve approach.

\subsection{CMS swaps}

A Constant Maturity Swap (CMS) is a swap with two legs of payments: on one side we receive (or, alternatively, we pay) in $T_i$, $i = 1,\dots,n$ the $c$-years IRS rate resetting in $T_{i-1}$, with $T_0=0$; on the other side, on the same payment dates, we pay (receive) the Libor rate corresponding to the period $\Delta$ going from $T_{i-1}$ to $T_i$ plus a spread $X_{n,c}$. The market quotes a value for $X_{n,c}$ which makes the swap fair; this fair spread can be expressed as follows:

\begin{equation}
\label{CMSswap}
X_{n,c} = \frac{\sum_{i=1}^n \left( \ExT{0}{T_i}{S_{T_{i-1}}(T_{i-1},T_c)} - {\tilde F}_0(T_i-\Delta,T_i) \right)P_0(T_i) }{\sum_{i=1}^n P_0(T_i)}
\end{equation}
where $T_c=T_{i-1}+c$ years and where all the accrual periods are considered to be equal to $\Delta$. In the Euro market $\Delta$ is equal to three-months, while the $c$-years IRS used as indexation in the CMS has Libor payments of six-months frequency.

Thus, CMS spreads theoretically depend on three different curves in our framework:
\begin{itemize}
\item a funding curve used to discount the cash flows of the CMS swap, which we consider to be the risk-free curve since CMS swaps are collateralized products; notice that this curve is also used as discounting curve when evaluating the indexation IRS rate payed by the CMS swap;
\item a three-months forwarding curve for the EURIBOR rates payed in the second leg of the CMS;
\item a six-months forwarding curve for the EURIBOR rates payed by the indexation IRS.
\end{itemize}

\begin{figure}
\begin{center}
\includegraphics[scale=0.5]{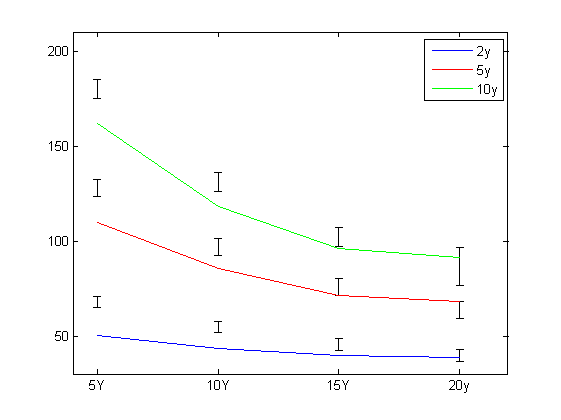}
\hfill
\includegraphics[scale=0.5]{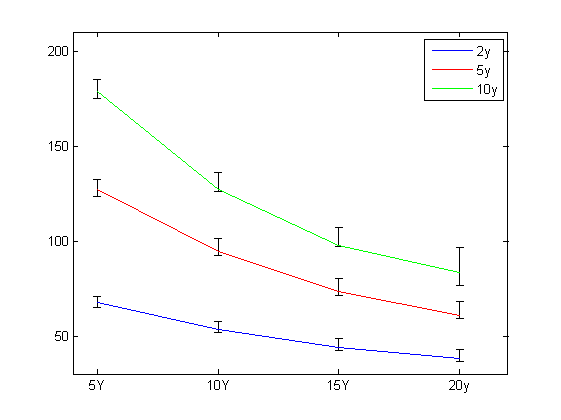}
\end{center}
\caption{
\label{fig:cmsswap}
Results of the calibration of the SABR model to swaption volatility smiles and CMS swaps spreads quoted by the market: left panel corresponds to the single-curve framework, right panel to the complete multi-curve framework. Continuous lines are CMS spreads implied by the model vs. underlying swap tenors, error bars are market bid-ask spreads. Different curves correspond to different swap tenors: two, five and ten years. $x$-axis is swap maturity, $y$-axis is the value of the CMS spread (in basis points). Market data observed on 14 June 2010.
}
\end{figure}

We evaluate the expectation of the swap rates under the discounting forward measure $\ExT{0}{T_i}{S_{T_{i-1}}(T_{i-1},T_c)}$, appearing in Equation \eqref{CMSswap}, by using the SABR model according to market practice. We follow Mercurio and Pallavicini (2006) and we introduce the function
\[
{\bar f}(x) := \frac{\left(1+\delta x\right)^{-\frac{T-T_a}{\delta}}}
                    {\sum_{i=a+1}^c \frac{T_i-T_{i-1}}{\left(1+(T_i-T_{i-1}) x\right)^{i-c}}} \ ,  
\]
where $\delta$ is the year fraction of the floating leg of the CMS swap (usually three-months for Euro area), and we calculate the relevant expectation by means of a static hedge as given by:
\begin{eqnarray}
\label{CMSswapprice}
\ExT{0}{T}{S_{T_a}(T_a,T_c)}
&=& \frac{{\bar f}(0)}{{\bar f}(S_0)} \, S_t(T_a,T_c) \\ \nonumber
& & + \frac{1}{{\bar f}(S_0)} \int_0^{+\infty} dx \, \left({\bar f}''(x)x + 2{\bar f}'(x)\right) {\rm Bl}(S_t(T_a,T_c),x,v(x,T_a))
\end{eqnarray}
where ${\rm Bl}(S,K,v)$ is the core of the Black formula for swaptions and the variance $v$ can be calculated starting from the SABR model. For instance, Hagan {\it et al.}~(2002) reports the following approximated variance
\[
v(K,T) := \sigma_{\rm SABR}^2(K,S) T
\]
with
\[
\begin{split}
\sigma_{\rm SABR}(K,S)\approx&
\dfrac{\alpha}{(SK)^\frac{1-\beta}{2}
\left[1+\frac{(1-\beta)^2}{24}\ln^2\!\left(\frac{S}{K}\right)+
\frac{(1-\beta)^4}{1920}\ln^4\!\left(\frac{S}{K}\right)\right]}
\frac{z}{x(z)}\\
&\cdot\left\{1+\left[\frac{(1-\beta)^2\alpha^2}
{24(SK)^{1-\beta}}+\frac{\rho\beta\epsilon\alpha}
{4(SK)^\frac{1-\beta}{2}}+\epsilon^2\frac{2-3\rho^2}{24}\right]T
\right\},
\end{split}
\]
where $\alpha$, $\beta$, $\rho$ and $\epsilon$ are SABR model parameters and
\[
z:=\frac{\epsilon}{\alpha}(SK)^\frac{1-\beta}{2}\ln\left(\frac{S}{K}\right)
\]
and
\[
x(z):=\ln\left\{\frac{\sqrt{1-2\rho
z+z^2}+z-\rho}{1-\rho}\right\}.
\]
but see also for better approximations Berestycki {\it et al.}~(2004), Ob\l\'oj (2008), Johnson and Nonas (2009), or Rebonato (2010).

We consider a different SABR model for each swap rate contained in the CMS payoff and we perform a calibration of all the SABR parameters (four parameters for each swap rate) to swaption volatility smile and CMS spreads quoted by the market. See Mercurio and Pallavicini (2006) for a detailed description of the calibration procedure. Here, in order to achieve a better fit on the market, we modify their calibration procedure to allow the $\beta$ parameter to depend on swap-rate tenor, namely we consider the same $\beta$ for swap rates with different fixing dates but with the same tenor.

In Figure \ref{fig:cmsswap} we show the calibration results: the former (left panel) is performed in the single-curve framework, while the second (right panel) in the multi-curve framework. Looking at the figure, it could seem that the complete multi-curve approach results in a better fit of market quotes. Actually, these results are not strong enough to make an assessment on the actual market practice; in fact, it would at least sound a little bit strange the fact that European swaptions are priced using a single-curve approach, while CMS swaps are priced using a multi-curve approach, since the convexity adjustments for the CMS rates are computed using swaption prices. Following these considerations, the CMS leg should be evaluated using a single-curve approach; on the other side, we cannot ignore the fact that the CMS is actually a swap, and the swap market works using multi-curves, quoting in particular a spread between the six-months and the three-months tenor rates.

\begin{figure}
\begin{center}
\includegraphics[scale=0.5]{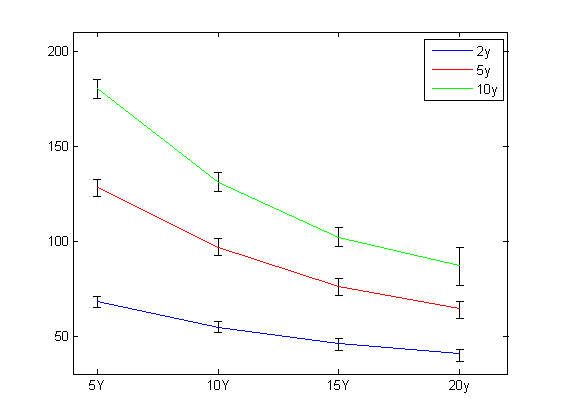}
\end{center}
\caption{
\label{fig:cmsswaph}
Results of the calibration of the SABR model to swaption volatility smiles and CMS swaps spreads quoted by the market: the panel corresponds to the ``hybrid" approach. Continuous lines are CMS spreads implied by the model vs. underlying swap tenors, error bars are market bid-ask spreads. Different curves correspond to different swap tenors: two, five and ten years. $x$-axis is swap maturity, $y$-axis is the value of the CMS spread (in basis points). Market data observed on 14 June 2010.
}
\end{figure}

Therefore, we could try and use an ``hybrid" setting, where we price the CMS leg using a six-months curve and the floating leg using a three-months curve. In Figure \ref{fig:cmsswaph} we report the results of the calibration resulting from this approach and we see that we obtain a very good fit of market quotes, comparable with the complete multi-curve calibration in the right panel of Figure \ref{fig:cmsswap}.

These results lead to a couple of important considerations: \textit{i}) the first key point is that it is fundamental to price the floating leg using its appropriate forwarding curve, because the existence of a spread between six-months and three-months rates is an important market fact that cannot be neglected; \textit{ii}) the CMS leg can be priced using both a single-curve or a multi-curve approach, obtaining similar results: this is due to the fact that the differences in the curves are recovered by the calibration of the SABR model, which returns different calibration parameters in each different framework. Notice, in particular, that in all cases the swaption volatility smiles are very well recovered by the calibrated SABR models.

%\begin{remark} {\bf Market practice to price CMS swaps:}
%A standard market practice to price CMS swaps is to evaluate products, depending on different underlying-rate tenors in a linear way such as CMS swaps, by splitting them into two instruments each depending on only one underlying-rate tenor, and then by using a single-curve framework to price each instrument, although such practice is not consistent for interest-rate theory, since different discounting curves are used within the same instrument.

%For instance, for CMS swaps the market practice is to evaluate the leg paying the IRS rate by using, for discounting and forwarding, a unique curve which is obtained only from six-months products, while the leg receiving the EURIBOR rate is priced by using again a unique curve for discounting and forwarding, but obtained only from three-months products.

%If we employ such scheme for CMS swap pricing we obtain a good agreement with market quotes, so that it is difficult to discriminate if the market has really shifted to a true multi-curve pricing framework.
%\end{remark}\label{rem:cms}

\subsection{CMS spread options}

The market quotes\footnote{Usually in the market we can find prices of strip of options, i.e. prices of cap/floor options on the spread between two swap rates. Also, it is possible to find prices for single caplets at different expiries. For sake of simplicity, in this section we deal with this second type of options.} the price $\pi_0(T_a,T_b,T_c;K)$ of an option with strike rate $K$ paying in $T_a$ the spread between two swap rates both fixing in $T_a$, the first maturing in $T_b$ and the second in $T_c$. The option price can be expressed as follows:

\begin{equation}
\label{CMSspread}
\pi_0(T_a,T_b,T_c;K) = P_0(T_a)\,\ExT{0}{T_a}{\left(S_{T_a}(T_a,T_b)-S_{T_a}(T_a,T_c)-K\right)^+}
\end{equation}
where the rates are accrued on a unitary period.

By following market practice we consider the two swap rates as jointly lognormally distributed with correlation $\rho_{bc}$. Thus, we can evaluate the spread option price by means of a generalization of the Margrabe formula, see for instance Brigo and Mercurio (2006).
\begin{equation}
\label{CMSspreadprice}
\pi_0(T_a,T_b,T_c;K) = P_0(T_a) \frac{1}{\sqrt{2\pi}} \int_{-\infty}^{+\infty} e^{-\frac{1}{2}x^2} \,{\rm Bl}(f(x),k(x),u) \,dx
\end{equation}
where the effective forward rate $f$, strike rate $k$ and variance $u$ are given by
\[
f(x) := \ExT{0}{T_a}{S_{T_a}(T_a,T_b)} \exp\left\{-\frac{1}{2}\rho_{bc}^2\sigma_b^2T_a + \rho_{bc}\sigma_b\sqrt{T_a}\,x\right\}
\]
\[
k(x) := K + \ExT{0}{T_a}{S_{T_a}(T_a,T_c)} \exp\left\{-\frac{1}{2}\rho_{bc}^2\sigma_c^2T_a + \sigma_c\sqrt{T_a}\,x\right\}
\]
\[
u := \sigma_b^2(1-\rho_{bc}^2)T_a
\]
being the expectations calculated by means of the SABR model previously calibrated to the swaption volatility smiles and to the CMS spreads (see the previous section), and the volatilities $\sigma_a$ and $\sigma_b$ are taken from swaptions market quotes by interpolating the volatility surface respectively at the following strikes:
\[
K_a := K + \ExT{0}{T_a}{S_{T_a}(T_a,T_c)}
\;,\quad
K_b := \left( \ExT{0}{T_a}{S_{T_a}(T_a,T_c)} - K \right)^+
\]

\begin{figure}
\begin{center}
\includegraphics[scale=0.5]{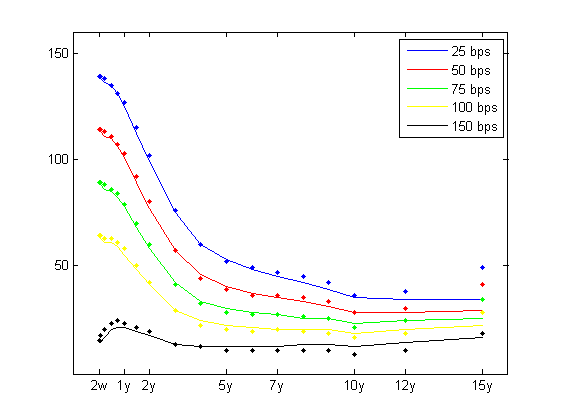}
\hfill
\includegraphics[scale=0.5]{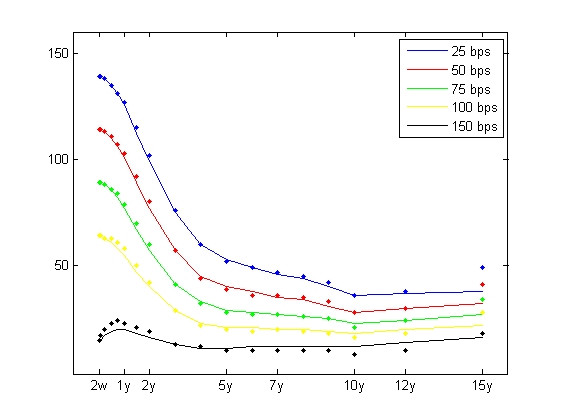}
\end{center}
\caption{
\label{fig:cmsspread_flat}
Calibration of spread option prices quoted by the market with a flat correlation: the left panel corresponds to the single-curve framework, the right panel corresponds to the multi-curve framework. Continuous lines are model prices, dots are market quotes. Different curves correspond to different option strikes ranging from 25 to 150 bps. $x$-axis is the option expiry, $y$-axis is the option price (in basis points). Market data observed on 14 June 2010.
}
\end{figure}

The only unknown parameter is the correlation $\rho_{bc}$, which can be calibrated to quoted spread option prices; in particular, in the market several quotes corresponding to different strike rates $K$ are available, and we use a unique flat correlation value for each strike~$K$. The results of the calibration are shown in Figure \ref{fig:cmsspread_flat}. We notice that the multi-curve approach (right panel) does not present a significantly better fit with respect to the single-curve approach (left panel). This fact can be explained considering that for CMS spread options we haven't a second leg paying a rate with different tenor like in the case of CMS swaps, so there is not a real need for multi-curve pricing; also, as we have previously seen, the SABR model is flexible enough to fit market data when calibrated either in single-curve or in multi-curve setting.

\subsection{Review of the results}

Let us summarize all the results presented in the previous sections.
\begin{itemize}
\item \textbf{IRS}: they seem to be priced using a multi-curve framework; in particular, market practice seems to be coherent with a discounting curve based on overnight rates. Notice that coherence with spot-starting swaps is granted by construction of the indexation curves; on the other side, this is not granted for forward starting swaps. The goodness of the fit we have found in our analysis is a strong point in favour of the multi-curve framework.
\item \textbf{Swaptions}: they seem to be priced using a classical single-curve framework based on a six-months curve. Notice, then, a market segmentation between the swap sector and the option sector.
\item \textbf{CMS}: the fit of CMS spreads cannot be considered a strong proof of the use of a complete multi-curve pricing. As we have seen, we just can state that CMS cannot be priced in a single-curve framework, since we cannot ignore the basis spread between the six-months and the three-months rates. Even if the hybrid approach previously presented is probably the actual market practice coherent both with swap and swaption pricing, also the complete multi-curve approach can be used, leaving to the SABR model parameters the task to fit market quotes.
\item \textbf{CMS spread options}: as for the plain-vanilla CMS, they cannot be used as an evidence of a multi-curve pricing, since all the dynamics parameters can be ``hidden" in the correlation structure or in the SABR calibrated parameters.
\end{itemize}

Thus, it seems that the market has moved to a multi-curve setting for what concerns the pricing of plain-vanilla instruments like IRS, but the situation is not so clear for derivative contracts, where the calibration of volatility and correlation parameters may hide the impact of which yield curve is used in pricing. In particular, this holds for CMS swaps and CMS options, while the swaption market has a strong evidence of pricing in the old single-curve approach.
%, where the complete multi-curve approach (that is two curves, one for discounting and one for indexation) returns comparable results with the single-curve approach (one unique curve used both for discounting and indexation). Probably it is due to the fact that a strong model for derivatives in the multi-curve framework is still missing.

\section{A minimal model formulation within the multi-curve HJM framework}
\label{sec:gaussian}

The tempative to introduce a multi-curve interest-rate model able to reproduce all the market quotes must face the problem of recovering prices which are not always coherent with a multi-curve approach, as we have seen in the previous section.

Further, the market does not quote options on all rate tenors. In the Euro area only options on the six-months tenor are widely listed, while the three-months tenor is present only in few quotes (swaptions with one-year tenor and cap/floors with maturities up to two years), and options on the other rate tenors are missing. Thus, any model which requires a different dynamics for each term-structure, has the problem that market quotes cannot be found to fix all its degrees of freedom.

Here, we select a simple but realistic volatility specification for the multi-curve HJM framework presented in the second section, which is able to fit the swaption prices and CMS swap spreads quoted by the market. Then, an exotic option pricing could be performed by Monte Carlo simulations.

In particular, we define a multi-curve model with a unique common dynamics between all the tenors, by extending the uncertain parameters Gaussian models, presented in Mercurio and Pallavicini (2005) within the single-curve framework. We address to further works the more ambitious task to define a fully stochastic volatility model within the HJM multi-curve framework.

%Once the discounting and the forwarding curves are known, we can calibrate our HJM model to quoted interest-rate options by keeping as constraints the zero-coupon bonds calculated from these curves. 

%The analysis of market data reported in the previous sections has shown a direct evidence of derivative pricing by means of a multi-curve framework. In particular forward interest-rate swaps, CMS swaps and CMS spread options agree with a different choice for term-structures for discounting and forwarding. Notice also that to evaluate CMS swaps we are forced to use three different curves since such instruments depend on two different EURIBOR tenors.

\subsection{Gaussian models with uncertain parameters}

We select the instantaneous forward volatilities of the HJM model in order to obtain a Gaussian model with uncertain parameters as in Mercurio and Pallavicini (2005). First, we consider that the instantaneous forward volatilities are all equal and all the movements of all term structures are perfectly correlated to the movements of the discount curve. Then, we define
\[
\sigma^\Delta_t(T)\,dW^{\Delta}_t := \sum_{k=1}^{q(I)} \sigma_k(t;I) \,e^{-a_k(I)(T-t)} \,dW_k(t)
\]
where $I$ is a discrete random variable, independent of the Brownian motions, taking values in the set $\{1,\ldots,m\}$ with probabilities $\omega_i:=\Qx{I=i}>0$ and $\sum_{i=1}^m \omega_i=1$, while $\sigma(t;I)$ are positive deterministic functions and $a(I)$ are positive constants. Further, in each scenario $i$ we allow the model to be driven by $q(i)$ Brownian motions $W_{1},\ldots,W_{q(i)}$, whose number may be different varying the scenarios. The Brownian motions are correlated according to
\[
\rho_{kh} := \frac{d}{dt}\langle W_k,W_h\rangle_t
\]

With such choice we obtain the following Minimal Multi-Curve Gaussian (MMG) model under discounting risk-neutral measure:
\begin{equation}
\label{eq:rUPM}
r = \varphi(t) + \sum_{k=1}^{q(I)} x_k(t;I)
\;,\quad
r^\Delta(t) = \varphi^\Delta(t) + \sum_{k=1}^{q(I)} x_k(t;I)
\end{equation}
with
\[
dx_k(t;I) = -a_k(I) x_k(t;I)\,dt + \sigma_k(t;I)\, dW_k(t),
\]
where $r$ is the short rate entering the definition of the discounting zero-coupon bond $P$, and $r^\Delta$ is the short rate entering the definition of the forwarding zero-coupon bond $P^\Delta$.

Notice that, to reproduce the initial yield curves, we introduce a deterministic function $\varphi^\Delta$ for each rate tenor and a deterministic function $\varphi$ for the discounting curve. An alternative approach may add uncertainty also on the $\varphi$s, see for instance the discussion in Mercurio and Pallavicini (2005).

All prices with MMG model can be calculated by explicitly taking the expectations on the discrete random variable $I$ since it is independent of the Brownian motions. Indeed, we get for a generic payoff $\Pi(t,T)$
\[
\Ex{t}{\Pi(t,T)} = \Ex{t}{\Ex{t}{\Pi(t,T)\big|I=i}} = \sum_{i=1}^m \omega_i \,\Ex{t}{\Pi(t,T)\big|I=i}
\]

\subsection{Numerical results}

We calibrate the MMG model to the following market quotes: at-the-money (ATM) swaption volatilities, swaption volatility smile and CMS spreads. %We get market quotes from ICAP\textsuperscript{\textregistered} pages listed on Bloomberg\textsuperscript{\textregistered} platform on 30 of March 2010.

%Notice that, when we refer to the multi-curve framework, we consider that the initial yield curves for the MMG model are calibrated as described in section \ref{sec:bootstrap}, while for the single-curve case we consider the same mixture of Gaussian models, but with a unique initial term structure which is bootstrapped by using the usual procedure which can be found for instance in Hagan and West (2006).

Model calibration is performed by minimizing the $L^2$-distance between market quotes and model-implied quotes. The calibration engine is the Levenberg-Marquardt algorithm implemented by Lourakis (2004).

\begin{figure}
\begin{center}
\includegraphics[scale=0.5]{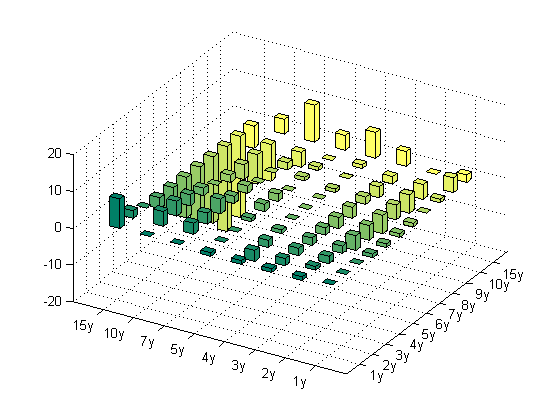}
\hfill
\includegraphics[scale=0.5]{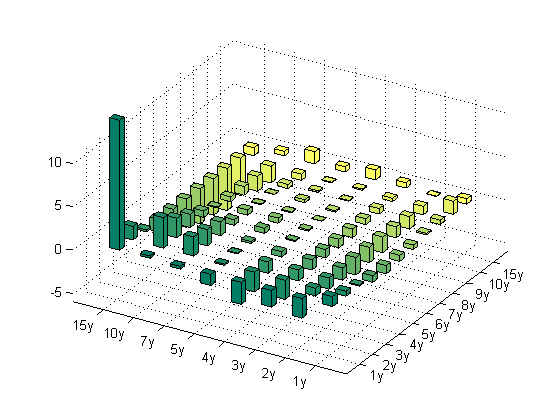}
\end{center}
\caption{
\label{fig:atmvol}
Swaption ATM prices, calculated with a two-scenario two-factor Gaussian model for different swap tenors (right axis) and expiries (left axis). Swaptions with one-year tenor are on three-months underlying rate, while the others are on six-months rate. Each bar represents the difference between market and calibrated prices (absolute errors in basis points on the left and relative errors in percentage on the right). Market data observed on 14 June 2010.
}
\end{figure}

\begin{figure}
\begin{center}
\includegraphics[scale=0.5]{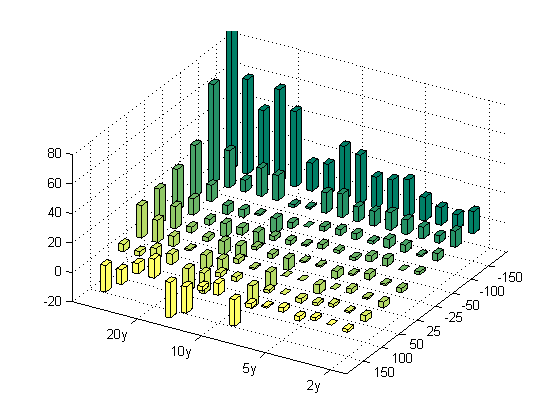}
\hfill
\includegraphics[scale=0.5]{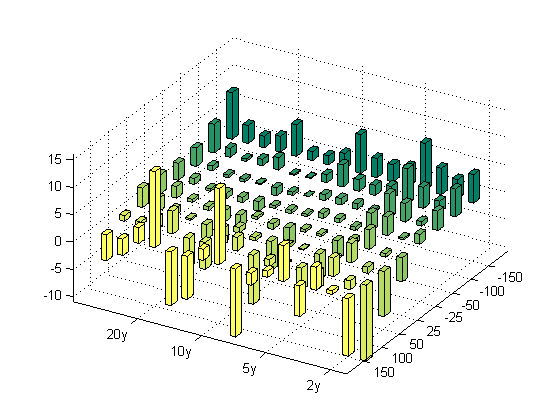}
\end{center}
\caption{
\label{fig:smile}
Swaption out-of-the-money prices, calculated with a two-scenario two-factor Gaussian model for different swap tenors and expiries (left axis) and differences (in basis points) between strike and ATM forward rate (right axis). On the left axis each swap tenor corresponds to four rows each referring to a different expiry (from right to left within the left axis they are 2y, 5y, 10y and 20y). Each bar represents the difference between market and calibrated prices (absolute errors in basis points on the left and relative errors in percentage on the right). Market data observed on 14 June 2010.
}
\end{figure}

In Figure \ref{fig:atmvol} we report the results of the calibration to ATM swaptions prices, while in Figure \ref{fig:smile} we report the results of the calibration to swaption smiles. %we analyze the swaption sector: we see that the results are quite similar between the single-curve and the multi-curve approaches, with the exception of the one-year tenor swaption. Indeed, in the Euro market, one-year tenor swaptions are quoted having as underlying an IRS paying a three-months EURIBOR, so that, as expected, a multi-curve framework performs better on such tenor, since a single-curve model cannot discriminate between swaptions quoted on different EURIBOR tenors.

%\begin{figure}[!htp]
%\begin{center}
%\includegraphics[scale=0.7]{figure/atmcap_scurve.png}
%\vfill
%\includegraphics[scale=0.7]{figure/atmcap_pcurve.png}
%\end{center}
%\caption{
%\label{fig:atmcap}
%At-the-money cap volatilities, calculated with a two-scenario two-factor Gaussian model for different cap maturities. Continuous lines are model implied volatilities, dots are market quotes. Cap up to two-years maturity are on three-months underlying rate, while cap longer than two years are on six-months rate. $x$-axis is cap maturity, $y$-axis is at-the-money cap volatility. Upper panel is single-curve framework, lower panel is multi-curve framework.
%}
%\end{figure}
%
%In Figures \ref{fig:atmcap} we analyze the cap/floor sector: we see that the results are quite different on cap maturities under three years, since these quotes refer to cap over three-months EURIBOR. As already noticed in the swaption sector, we see that, when the model must discriminate between forward rate tenors, it should include in a coherent way all the yield curves, and in fact we have a much better fit in the multi-curve framework with respect to the single-curve framework.

\begin{figure}
\begin{center}
\includegraphics[scale=0.55]{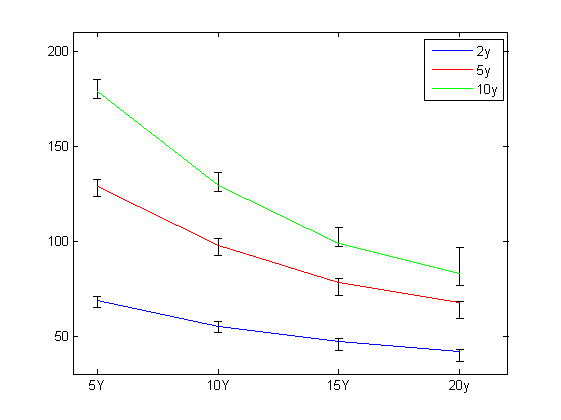}
\end{center}
\caption{
\label{fig:cms}
CMS spreads, calculated with a two-scenario two-factor Gaussian model for different maturities. Continuous lines are CMS spreads implied by the model vs. underlying swap tenors, error bars are market bid-ask spreads. Different curves correspond to different swap tenors: two, five and ten years. $x$-axis is swap maturity, $y$-axis is the value of the CMS spread (in basis points). Market data observed on 14 June 2010.
}
\end{figure}

Finally, in Figure \ref{fig:cms} we analyze the CMS sector: we see that the results are quite similar to the ones obtained with the SABR model in section \ref{sec:evidence}. As already discussed earlier, the good fit observed is coherent with the use of a three-months based curve for the EURIBOR leg.

Globally, we have found quite good calibration results: the larger errors are in the extreme wings of the swaption smiles, but this can be explained by the fact that the Gaussian mixture is by construction less flexible than the SABR model to accomodate smiles. Neverthless, the results are good enough to let the calibrated model to be used for Monte Carlo simulations when pricing more complicated exotic instruments.

\section{Conclusions}
\label{sec:conclusions}

Recent turmoils in the world economies and in particular in the financial markets have shown that credit and liquidity issues are crucial when evaluating financial products. Fundamental assumptions used for many years in the interest rate market need to be revised: the concept of \textit{risk-free} yield curve is different from the standard idea of discounting curve used in the past, and all the models used for pricing must be improved to take into account these new features.

In this paper we showed that it is fundamental to think of rates with different tenors as if they were different assets, since the credit risk and the liquidity problems make them actually different contracts. We have followed a HJM approach to define both the dynamics of the discounting curve and the dynamics of the yield curves used to calculate forward rates with different tenors.

We presented a methodology which allows to bootstrap market quotes of plain-vanilla interest rate instruments in order to obtain a set of initial forwarding term structures, one for each rate tenor, and we illustrated how this approach works in practice. In particular we analyzed the case of forward starting IRS, plain-vanilla swaptions, CMS swaps and CMS spread options, showing that the market has not moved yet to a complete multi-curve pricing framework. On one side it is fundamental to price swap contracts including rates of different tenors by means of the multi-curve setting, in order to be coherent with the non-null quoted basis swap spreads; on the other side, swaptions are still requiring a single-curve approach, while CMS derivatives, once the three-month leg is consistently priced, can be values either using a single-curve or a multi-curve approach.

Finally, we presented a simple extension of a mixture of Gaussian models (MMG model) incorporating the multi-curve structure: we showed that it is possible to obtain a coherent calibration to a various set of options including basic rates with different tenors. The calibrated model can then be used in a Monte Carlo simulation when pricing more complicated exotic options.

The multi-curve HJM framework we proposed can be easily extended to incorporate different dynamics for each yield curve, possibly with stochastic volatilities, so that, if the market starts to quote a significant amount of options on different interest-rate tenors or even on basis-swap spreads, we could include them in the calibration of models based on our framework.


\begin{thebibliography}{99}

\bibitem {Ametrano09}
F. Ametrano, M. Bianchetti (2009). Bootstrapping the Illiquidity: Multiple Yield Curves Construction For Market Coherent Forward Rates Estimation. Published in ``Modeling Interest Rates: Latest Advances for Derivatives Pricing", edited by F.~Mercurio, Risk Books.

\bibitem {Berestycki04}
H. Berestycki, J. Busca and I. Florent (2004), Computing the implied volatility in
Stochastic Volatility Models, Communications on Pure and Applied Mathematics, Vol. 57, No. 10, 1352–1373, October.

\bibitem {Bianchetti09}
M. Bianchetti (2009). Two Curves, One Price: Pricing ad Hedging Interest Rate Derivatives Using Different Yield Curves for Discounting and Forwarding. Available at {\tt http://ssrn.com/abstract=1334356}.

\bibitem {Bianchetti10}
M. Bianchetti (2010). Multiple Curves, One Price: The Post Credit-Crunch Interest Rate Market. Talk kept at ``Risk and modeling fixed income interest rates", Marcus Evans conference, London, 15-16 April.

\bibitem {Boenkost05}
W. Boenkost and W.M. Schmidt (2005). Cross currency swap valuation. Available at {\tt http://ssrn.com/abstract=1375540}.

\bibitem{BrigoCapponi}
D. Brigo, and A. Capponi (2008). Bilateral counterparty risk valuation
with stochastic dynamical models and application to Credit Default Swaps. Available at \\
{\tt http://ssrn.com/abstract=1318024} or at {\tt http://arxiv.org/abs/0812.3705}.

\bibitem {Brigo06}
D. Brigo, and F. Mercurio (2006). Interest Rate Models: Theory and Practice -
with Smile, Inflation and Credit, Second Edition, Springer Verlag.

\bibitem {Chibane09}
M. Chibane and G. Sheldon (2009). Building Curves on a Good Basis. Available at {\tt http://ssrn.com/abstract=1394267}.

\bibitem {Fruchard95} E. Fruchard, C. Zammouri and E. Willems (1995). Basis for change, Risk, Vol. 8, No.10 , 70-75, October.

\bibitem {Fujii10} 
M. Fujii, Y. Shimada and A. Takahashi (2010). On the Term Structure of Interest Rates with Basis Spreads, Collateral and Multiple Currencies. Available at {\tt http://ssrn.com/abstract=1556487}

\bibitem {Hagan02}
P. S. Hagan, D. Kumar, A.S. Lesniewski, and D.E. Woodward (2002). Managing Smile Risk. Wilmott magazine, September, 84-108.

\bibitem {Hagan06}
P. S. Hagan and G. West (2006). Interpolation Methods for Curve Construction. Applied Mathematical Finance, Vol. 13, No. 2, 89-129, June 2006

\bibitem {Henrard07}
M. Henrard, M. (2007). The Irony in the Derivatives Discounting. Wilmott Magazine, July 2007, 92-98.

\bibitem {Henrard09}
M. Henrard (2009). The Irony in the Derivatives Discounting Part II: The Crisis. Preprint, Dexia Bank, Brussels.

\bibitem {Johnson09}
S. Johnson and B. Nonas (2009). Arbitrage-free construction of the swaption cube. Available at {\tt http://ssrn.com/abstract=1330869}

\bibitem {Kenyon10}
C. Kenyon (2010). Short-Rate Pricing after the Liquidity and Credit Shocks: Including the Basis. Available at {\tt http://ssrn.com/abstract=1558429}

\bibitem {Kijima09}
M. Kijima, K. Tanaka and T. Wong (2009). A Multi-Quality Model of Interest Rates, Quantitative Finance 9(2), 133-145.

\bibitem {Liu08}
D. Liu D. and M. Wu (2008)
Inflation Modeling. Available at \\{\tt http://ssrn.com/abstract=1286502}

\bibitem {Lourakis04}
M. Lourakis (2004). A Brief Description of the Levenberg-Marquardt Algorithm Implemented by {\tt levmar}. Available at\\
{\tt http://www.ics.forth.gr/$\sim$lourakis/levmar/levmar.pdf}

\bibitem {Mercurio05}
F. Mercurio and A. Pallavicini (2005). Mixing Gaussian Models to Price CMS Derivatives. Available at {\tt  http://ssrn.com/abstract=872708}

\bibitem {Mercurio06}
F. Mercurio and A. Pallavicini (2006). Smiling at Convexity. Risk, August, 64-69.
An extended version is available at {\tt http://ssrn.com/abstract=892287}

\bibitem {Mercurio09}
F. Mercurio (2009). Interest Rates and The Credit Crunch: New Formulas and Market Models. Bloomberg Portfolio Research Paper No. 2010-01-FRONTIERS. Available at {\tt http://ssrn.com/abstract=1332205}

\bibitem {Mercurio10}
F. Mercurio (2010). LIBOR Market Models with Stochastic Basis. Available at {\tt http://ssrn.com/abstract=1563685}

\bibitem {Morini09}
M. Morini (2009). Solving the Puzzle in the Interest Rate Market. Available at {\tt http://ssrn.com/abstract=1506046}

\bibitem {Morini10}
M. Morini and A. Prampolini (2010). Risky Funding: A unified framework for counterparty and liquidity charges. Available at {\tt http://www.defaultrisk.com}

\bibitem {Obloj08}
J. Ob\l\'oj (2008). Fine-tune your smile correction to Hagan et al, Wilmott magazine, May/June 2008 p102.

\bibitem {Pallavicini10}
A. Pallavicini (2010). Counterparty Risk Evaluation for Interest Rate Derivatives. Talk kept at ``Risk and modeling fixed income interest rates", Marcus Evans conference, London, 15-16 April.

\bibitem {Piterbarg10}
V. Piterbarg (2010). Funding beyond discounting: collateral agreements and derivatives pricing. Risk, February, 97-102.

\bibitem {Rebonato10}
R. Rebonato (2010). Approximate Solutions for the SABR model: improving the Hagan expansion. Talk kept at Global Derivatives, Paris, 18-20 May.

\end{thebibliography}
\end{document}